\documentclass[aip,cha,reprint,amsmath,amssymb,floatfix]{revtex4-2}
\usepackage{graphicx}
\usepackage{dcolumn}
\usepackage{bm}

\usepackage[utf8]{inputenc}
\usepackage[T1]{fontenc}
\usepackage{mathptmx}
\usepackage{xcolor}
\usepackage{physics}

\begin{document}

\title{Interlayer Synchronisation of Time-Varying Multiplex
Kuramoto--Sakaguchi Networks in the Chimera Regime}

\author{Muhittin Cenk Eser}
\affiliation{Department of Mathematics and Computer Science, Freie Universit\"at Berlin, Arnimallee 6, 14195 Berlin, Germany}

\author{Mustafa Riza}
\affiliation{Department of Physics, Eastern Mediterranean University, 99628 Famagusta, North Cyprus,via Mersin 10, Turkey}

\date{\today}

\begin{abstract}
We study interlayer synchronisation in a duplex network of $N=300$ nonlocally
coupled Kuramoto--Sakaguchi oscillators, with each layer operating in the
chimera regime.  The interlayer coupling is weak ($\sigma_{12}=0.01$), sparse,
and time-varying: a fixed number $N_{IL}$ of replica-node pairs are coupled
symmetrically, and the active links are randomly redistributed every $T_{swt}$
time units.  We characterise synchronisation by the time-averaged interlayer
order parameter $Z$, the master stability function $\Psi(\sigma_{12},T_{swt})$,
and the finite-time transverse Lyapunov exponent $\lambda_\perp$.  In the
static case, full synchronisation ($Z=1$, $\Psi<0$) requires all-to-all
interlayer coupling ($N_{IL}=N$).  Under temporal switching with
$T_{swt}\leq 25$, near-complete synchronisation is achieved with as few as
$N_{IL}\approx N/3$ links, while the intralayer chimera structure is preserved.
The master stability function confirms that short switching periods render the
transverse dynamics stable at link densities where static coupling fails, and
the transverse Lyapunov exponent heatmap delineates the critical link number as
a joint function of $N_{IL}$ and $T_{swt}$.  These results demonstrate that
temporal redistribution of sparse interlayer connections can stabilise
replica-node coherence in networks with spatially heterogeneous intralayer
dynamics.
\end{abstract}

\maketitle

\section{Introduction}

Synchronisation is one of the most fundamental forms of collective behaviour in
nonlinear science. It describes the spontaneous adjustment of rhythms in coupled
dynamical units and appears in physical, chemical, biological, technological, and
social systems, including Josephson-junction arrays, chemical oscillators, cardiac
and neural rhythms, laser arrays, power-grid dynamics, circadian systems, and
collective human activity~\cite{winfree1967,pikovsky2001,acebron2005,strogatz2000}.
In its modern form, synchronisation theory no longer concerns only whether a set
of oscillators can lock to a common rhythm. It asks how coherence depends on the
interaction function, the distribution of natural frequencies, the coupling
strength, network topology, temporal modulation, and the presence of multiple
interacting layers~\cite{arenas2008,boccaletti2014,rodrigues2016}. This network
perspective is particularly important because many real systems are not adequately
represented by a single graph. Neuronal, ecological, social, transportation, and
infrastructure systems often contain several interaction channels, and the same
unit may have replicas or counterparts in multiple layers. Such systems are more
naturally represented as multilayer or multiplex
networks~\cite{boccaletti2014,kivela2014}.

A distinctive phenomenon in multiplex networks is interlayer synchronisation, in
which each unit locks to its replica in another layer even though units within a
single layer need not be mutually synchronous. This must be distinguished from
intralayer synchronisation, the coherence of nodes within one layer: a multiplex
can exhibit replica-node synchronisation without complete synchronisation inside
each layer, which is exactly what makes interlayer synchronisation a useful probe
of structured spatial patterns~\cite{gambuzza2015,sevilla2016,delgenio2017,nicosia2017}.
For identical layers the interlayer synchronous manifold is
\begin{equation}
\theta^{(1)}_i(t) = \theta^{(2)}_i(t), \qquad i = 1, \dots, N,
\label{eq:manifold}
\end{equation}
or, more generally, by equality of the state vectors of corresponding replica nodes.
The stability of this manifold determines whether the two layers can evolve as copies
of one another while retaining whatever coherent, incoherent, clustered, or
chimera-like organisation exists within each layer.

\begin{figure}
\includegraphics{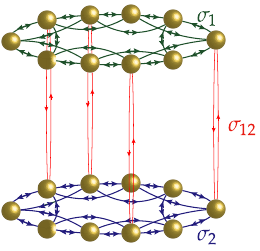}
\caption{Schematic of the two-layer Kuramoto--Sakaguchi multiplex defined by
Eqs.~\eqref{eq:model1}--\eqref{eq:model2}.  Each layer is a ring of $N$
oscillators with nonlocal coupling.  A subset of $N_{IL}$ replica-node pairs are
connected by interlayer links, which are randomly redistributed every $T_{swt}$
time units.}
\label{fig:model}
\end{figure}

The present study considers two identical layers of nonlocally coupled
Kuramoto--Sakaguchi oscillators,
\begin{align}
\dot{\theta}^{(1)}_i(t) &= \omega_i
  - \frac{\sigma}{2R}\sum_{j=i-R}^{i+R}
    \sin\!\big(\theta^{(1)}_i - \theta^{(1)}_j + \alpha\big) \nonumber\\
  &\quad + \chi_i(t)\,\sigma_{12}
    \sin\!\big(\theta^{(2)}_i - \theta^{(1)}_i + \alpha_{12}\big), \label{eq:model1}\\[2pt]
\dot{\theta}^{(2)}_i(t) &= \omega_i
  - \frac{\sigma}{2R}\sum_{j=i-R}^{i+R}
    \sin\!\big(\theta^{(2)}_i - \theta^{(2)}_j + \alpha\big) \nonumber\\
  &\quad + \chi_i(t)\,\sigma_{12}
    \sin\!\big(\theta^{(1)}_i - \theta^{(2)}_i + \alpha_{12}\big), \label{eq:model2}
\end{align}
with identical intralayer parameters $\sigma = 0.1$, $\alpha = 1.47$,
$\omega_i = 0$, and coupling radius $R$. Indices are understood modulo $N$, so
that each layer is a periodic ring; throughout this work we use $N=300$ and a
coupling radius $R=105$, corresponding to a coupling-range fraction
$r \equiv R/N = 0.35$, which places each isolated layer firmly in the
single-headed chimera regime for $\alpha=1.47$. The parameter $\alpha_{12}$ is an
optional interlayer phase lag; throughout this work we take $\alpha_{12} = 0$,
so that the interlayer coupling is purely attractive, and we retain $\alpha_{12}$
in the analytical expressions only to make the dependence on the interlayer phase
response explicit. The two layers are coupled symmetrically with weak interlayer
coupling strength $\sigma_{12} = 0.01$. The switching variable
$\chi_i(t) \in \{0,1\}$ indicates whether the replica pair $(i,i)$ is connected at
time $t$; temporal switching changes the locations of the active interlayer links
while keeping their number fixed. Interlayer synchronisation is therefore promoted
not by increasing the number of interlayer links, but by redistributing a fixed
number of symmetric interlayer connections in time. The problem therefore sits at the
intersection of three bodies of work: synchronisation in multilayer networks,
time-varying or blinking coupling, and chimera states in nonlocally coupled phase
oscillators. Figure \ref{fig:model} illustrates the setup schematically.

This architecture also abstracts a concrete biological system. The two cerebral
hemispheres form two near-identical layers of nonlocally coupled neural
populations, with homologous left--right regions as replica pairs $(i,i)$ and the
callosal projections as interlayer links~\cite{anesiadis2022}. Here the chimera
regime is the physiologically relevant state: in unihemispheric sleep---documented
in birds and aquatic mammals---one hemisphere stays coherent (awake) while the
other is desynchronised (asleep), so a faithful model must lock the hemispheres as
replicas \emph{without} erasing the partial coherence within
each~\cite{ramlow2019}. Because interhemispheric coupling is metabolically costly
and dynamically gated rather than static and all-to-all, a sparse, temporally
redistributed link set---a form of dynamic functional connectivity---is the more
realistic abstraction. The same picture connects to pathology: the abrupt collapse
of a chimera into global coherence has been proposed as a dynamical analogue of
epileptic-seizure onset, making the stability of the partially coherent manifold,
not mere synchronisability, the meaningful quantity~\cite{andrzejak2016}.
\subsection{Diagnostics}

Interlayer synchronisation is quantified by the interlayer order parameter
\begin{equation}
Z(t) = \left| \frac{1}{N} \sum_{j=1}^{N}
  \exp\!\Big[ i\big(\theta^{(1)}_j(t) - \theta^{(2)}_j(t)\big) \Big] \right|.
\label{eq:R12}
\end{equation}
$Z(t)=1$ whenever the replica phase differences are equal---in particular on the
manifold of Eq.~\eqref{eq:manifold}---and $Z(t)\approx 0$ for incoherently
distributed differences. Because the switched system shows long transients and
intermittent excursions toward synchrony, we use the time-averaged value
\begin{equation}
Z = \frac{1}{T_2 - T_1} \int_{T_1}^{T_2} Z(t)\, dt,
\label{eq:R12bar}
\end{equation}
over a window taken after transients. $Z$ measures replica-node phase locking
without requiring intralayer coherence, which makes it the natural diagnostic in
the chimera regime.

A large $Z$ shows that replicas are aligned but not that the alignment is
dynamically stable. Stability is decided by the response of the
manifold~\eqref{eq:manifold} to transverse perturbations
$\eta_i=\theta^{(1)}_i-\theta^{(2)}_i$, whose linearised evolution defines the
transverse Lyapunov exponent $\lambda_\perp$; $\lambda_\perp<0$ means the manifold
is attracting. This is the master-stability-function (MSF) viewpoint of Pecora and
Carroll~\cite{pecora1990,pecora1998}, extended to multiplex layers by Tang
\emph{et al.}~\cite{tang2019} and others~\cite{sevilla2016,delgenio2017,dellarossa2020}.
One subtlety is decisive here. The standard MSF reduction assumes a synchronous
trajectory \emph{common to all nodes}, so that the variational equation
block-diagonalises through the coupling-matrix spectrum. On the interlayer
manifold, however, each replica pair follows a node-dependent chimeric trajectory
$s_i(t)$; the Jacobian is then not co-diagonalisable with a single Laplacian, and
the eigenmode reduction of Ref.~\cite{tang2019} does not apply. We therefore obtain
$\lambda_\perp$ by integrating the full transverse system numerically
(Secs.~\ref{sec:MSF}--\ref{sec:TLE}). Since finite-$N$ chimeras are long chaotic
transients~\cite{wolfrum2011}, we evaluate a \emph{finite-time} $\lambda_\perp$
over the same window used for $Z$, so that both diagnostics judge the same observed
state. A large $Z$ together with $\lambda_\perp<0$ then certifies genuine,
attracting interlayer synchronisation.

\subsection{Background and the central question}

The intralayer dynamics is of Kuramoto--Sakaguchi type, the canonical
phase-reduction model of collective synchronisation across physics, biology,
neuroscience, and power grids~\cite{acebron2005,kuramoto1984,strogatz2000,rodrigues2016,dorfler2013,dorfler2014}.
The phase lag $\alpha$ introduced by Sakaguchi~\cite{sakaguchi1986} brings
frustration, and in a nonlocally coupled ring $\alpha$ and the coupling radius
jointly control the transition between coherence, incoherence, multichimera, and
chimera states~\cite{kuramoto2002,abrams2004,abrams2006,omelchenko2011,omelchenko2013,omelchenko2012,kemeth2016}.
Chimera states---the coexistence of coherent and incoherent domains among
identical, symmetrically coupled units---were identified by Kuramoto and
Battogtokh~\cite{kuramoto2002} and Abrams and
Strogatz~\cite{abrams2004,abrams2006,abrams2008}, rendered analytically tractable
by the Ott--Antonsen ansatz~\cite{ott2008,laing2009}, observed in optical,
chemical, and mechanical experiments~\cite{hagerstrom2012,tinsley2012,martens2013},
and surveyed in Refs.~\cite{panaggio2015,majhi2019}. With identical natural
frequencies $\omega_i=0$, our layers do not synchronise by overcoming frequency
disorder; the question is whether weak interlayer coupling can align two
already-structured chimera patterns without erasing them.

In a multiplex, interlayer coupling is not a passive addition to chimera dynamics:
it can shift incoherent domains, suppress or replicate a chimera, or lock two
chimeras to one another~\cite{ghoshjalan2016,majhi2017,sawicki2018,leyva2018,kumar2021,andrzejak2017}.
Maksimenko \emph{et al.}~\cite{maksimenko2016} showed that the same coupling can
either destroy a chimera or, when stronger, impose a shared ``interlayer
chimera''; this is precisely why a deliberately weak, sparse coupling is the
pertinent regime for locking the layers while preserving their internal structure.
Independently, time-varying or blinking couplings are known to stabilise synchrony
that no static snapshot supports, the fast-switching dynamics being governed by the
time-averaged coupling matrix~\cite{belykh2004,stilwell2006,rakshit2017,ghoshTVN2022,majhiTVN2022}.
The closest precedent, Buscarino \emph{et al.}~\cite{buscarino2015}, found that
time-varying interpopulation links generate and reshape chimeras in two Kuramoto
populations; adaptive multiplex couplings can likewise organise and synchronise
chimeras~\cite{kasatkin2017,kasatkin2018}.

These ingredients have not previously been combined. Classical theory explains
chimera formation and, through the multiplex MSF~\cite{tang2019}, interlayer
stability for \emph{homogeneous} intralayer states; what is missing is the regime
in which two chimera-bearing layers are joined by a fixed, sparse, time-switched
interlayer coupling, so that the synchronous trajectory is itself spatially
heterogeneous and the standard MSF reduction fails. The central question of this
paper is therefore: \emph{can temporal redistribution of a fixed, sparse set of
weak interlayer links render the interlayer-synchronous manifold attracting---without
increasing link density and without destroying the intralayer chimera?} We answer
it affirmatively. Using the three mutually consistent diagnostics introduced
above---$Z$, the MSF $\Psi(\sigma_{12},T_{swt})$, and $\lambda_\perp$---we show that
switching achieves near-complete interlayer synchronisation with about a third of
the links required statically, and we map the critical link number as a function of
the switching period.

\section{Time-Averaged Interlayer Order Parameter Analysis}
\label{sec:order}

Throughout this section the coupled layers~\eqref{eq:model1}--\eqref{eq:model2}
are integrated with a fixed-step fourth-order Runge--Kutta scheme
($\mathrm{d}t=0.01$) on the periodic ring of $N=300$ oscillators with $R=105$
($r=0.35$), $\sigma=0.1$, $\alpha=1.47$, $\omega_i=0$, $\sigma_{12}=0.01$, and
$\alpha_{12}=0$. After discarding an initial transient of $1000$ time units, the
instantaneous interlayer order parameter $Z(t)$ of Eq.~\eqref{eq:R12} is averaged
over the subsequent window $[T_1,T_2]$ to obtain $Z$ of Eq.~\eqref{eq:R12bar}; the
window is taken long enough to span many switching periods so that $Z$ is
insensitive to the precise endpoints. Unless stated otherwise, each reported $Z$
is additionally averaged over independent random realisations of the initial
phases and the interlayer-link configurations, as specified per figure.

\subsection{The static case}

\begin{figure}[!ht]
\centerline{\includegraphics[width=\columnwidth]{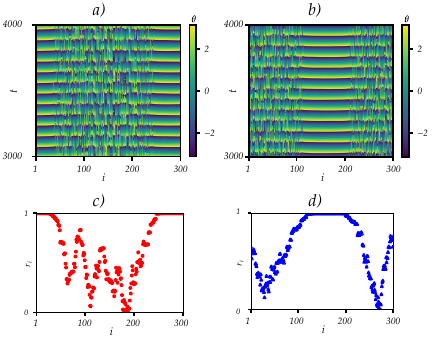}}
\caption{No interlayer connection ($N_{IL}=0$). Spacetime plots (phase colour-coded,
$\theta\in[-\pi,\pi]$) for layer~1 a) and layer~2 b), and local order parameters
$r_i=\big|\tfrac{1}{2\delta}\sum_{|j-i|\le \delta}e^{\mathrm{i}\theta_j}\big|$, where $\delta =10$ are the number of nearest neighbour nodes for the calculation of the local order parameter. for layer~1 c)
and layer~2 d) at $t=4000$ ($N=300$, $r=0.35$, $\sigma=0.1$, $\alpha=1.47$).
Each layer independently sustains a single-headed chimera---$r_i$ near unity in the
coherent domain, depressed in the incoherent one---with uncorrelated domain
locations, as expected without interlayer coupling.}
\label{fig:kuramoto_noilc}
\end{figure}

We begin by considering two uncoupled, identical layers of Kuramoto--Sakaguchi
oscillators in the chimera regime. The spacetime plots and local order parameters
$r_i$ for both layers at $t = 4000$, shown in Fig.~\ref{fig:kuramoto_noilc},
confirm that each layer independently sustains a chimera: a coherent spatial domain
with a smooth phase profile coexists with an incoherent domain with irregular phases.
The local order parameters of the two layers are uncorrelated, as expected in the
absence of any interlayer coupling.

\begin{figure}[!ht]
\centerline{\includegraphics[width=\columnwidth]{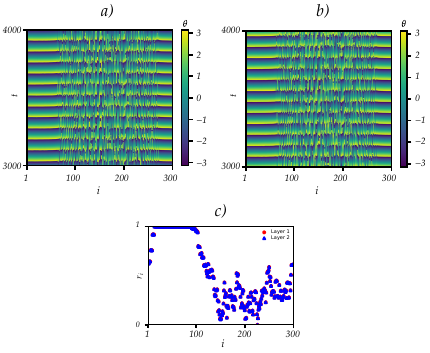}}
\caption{All-to-all interlayer connection ($N_{IL}=N=300$) with weak coupling
$\sigma_{12}=0.01\ll\sigma=0.1$. Spacetime plots for layer~1 a) and layer~2 b)
(phase colour-coded as in Fig.~\ref{fig:kuramoto_noilc}), and the local order
parameters 
$r_i=\big|\tfrac{1}{2\delta}\sum_{|j-i|\le \delta}e^{\mathrm{i}\theta_j}\big|$, where $\delta =10$ are the number of nearest neighbour nodes for the calculation of the local order parameter of both layers superimposed c). The coherent and incoherent domains
are now co-located and the two $r_i$ curves overlap exactly, confirming perfect
interlayer synchronisation ($Z=1$): even weak coupling locks the layers as
replicas while the shared chimera pattern persists.}
\label{fig:kuramoto_sync}
\end{figure}

When all replica nodes are connected to each other ($N_{IL} = N = 300$), even with
weak interlayer coupling ($\sigma_{12}=0.01 \ll \sigma=0.1$), the phases of each
node coincide perfectly with those of its replica. Figure~\ref{fig:kuramoto_sync}
shows that the coherent and incoherent spatial regions of both layers are
co-located, and the local order parameters of layers~1 and~2 overlap exactly
(panel~c)), confirming perfect interlayer synchronisation ($Z=1$).

\begin{figure}[!ht]
\centerline{\includegraphics[width=\columnwidth]{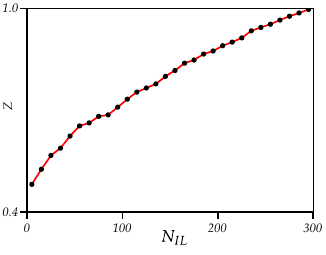}}
\caption{Time-averaged interlayer order parameter $Z$ as a function of $N_{IL}$
for the static (non-switching) case, averaged over 30 initial conditions.
Full synchronisation ($Z=1$) is reached only at $N_{IL}=N=300$.}
\label{fig:kur_interlayerorder_nosw}
\end{figure}

We next characterise synchronisation as a function of $N_{IL}$ in the static case,
averaging over 30 different initial conditions and 100 different interlayer
connection configurations. Figure~\ref{fig:kur_interlayerorder_nosw} shows that
$Z$ reaches~1 only when $N_{IL} = N = 300$, i.e.\ only when every replica pair is
connected. For any partial set of static links, the time-averaged order parameter
remains well below~1.

The ensemble mean of $Z$ at every value of $N_{IL}$ is already stable, within the
symbol size of Fig.~\ref{fig:kur_interlayerorder_nosw}, after roughly 20--30
realisations, and increasing the ensemble above 30 changes neither the
shape of the curve nor the location of any feature; 
On this basis, the switching sweeps---where each realisation is
substantially more expensive because the measurement window must cover many
switching periods---use 30 realisations, a number whose adequacy is verified
independently by the basin-of-attraction analysis of Fig.~\ref{fig:basin}, which
displays the full realisation-by-realisation scatter rather than only the mean.

\subsection{Time-varying interlayer links}

Following the methodology of Ref.~\cite{Eser21} for FitzHugh--Nagumo (FHN)
oscillator networks, we now investigate whether temporal switching of a fixed
number of interlayer links can achieve synchronisation with far fewer than
$N_{IL} = N$ connections. For a given $N_{IL}$, the active interlayer links are
randomly redistributed every $T_{swt}$ time units: after each interval of duration
$T_{swt}$, the current set of links is removed and a new set of $N_{IL}$ randomly
chosen replica pairs is connected. Statistically, successive connection patterns are
almost certainly different.

\begin{figure}[!ht]
\centerline{\includegraphics[width=\columnwidth]{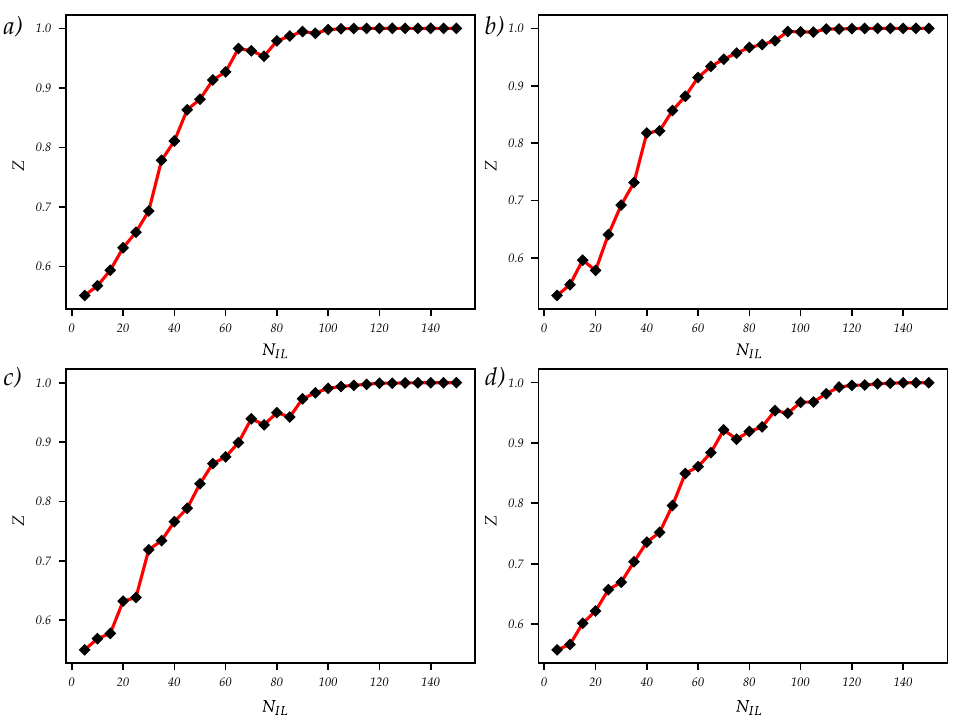}}
\caption{Time-averaged interlayer order parameter $Z$ as a function of $N_{IL}$
for switching times a)~$T_{swt}=25$, b)~$T_{swt}=50$, c)~$T_{swt}=75$,
and d)~$T_{swt}=100$.}
\label{fig:kur_interlayerorder}
\end{figure}

Figure~\ref{fig:kur_interlayerorder} shows $Z$ as a function of $N_{IL}$ for
switching times a)~$T_{swt}=25$, b)~$T_{swt}=50$, c)~$T_{swt}=75$, and
d)~$T_{swt}=100$. In all four cases, $Z$ increases monotonically with $N_{IL}$.
As $N_{IL}$ approaches 100, $Z$ approaches~1 in each panel, indicating
near-complete interlayer synchronisation across the full range of tested switching
times.

\begin{figure}
\centerline{\includegraphics[width=\columnwidth]{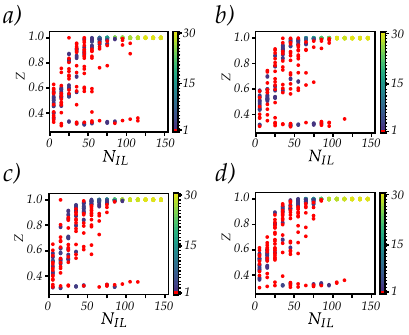}}
\caption{Basin-of-attraction analysis for switching times
(a)~$T_{swt}=25$, (b)~$T_{swt}=50$, (c)~$T_{swt}=75$, and
(d)~$T_{swt}=100$, over 30 network realisations. Dot colour indicates the
number of appearances: red corresponds to 1 appearance and yellow to 30.
For all four switching times, $Z=1$ is robustly achieved for
$N_{IL}\geq 100$.}
\label{fig:basin}
\end{figure}

To assess the robustness of this result, we analysed the basin of attraction across
30 independent network realisations for each of the four switching times.
Figure~\ref{fig:basin} shows the time-averaged order parameter for each
realisation, colour-coded by the number of times that value was observed (red for
one appearance, yellow for 30 appearances). For all four switching times, $Z$
reaches~1 for $N_{IL} \geq 100$, confirming that full interlayer synchronisation
is robust across different random connection sequences.

\begin{figure}
\centerline{\includegraphics[width=\columnwidth]{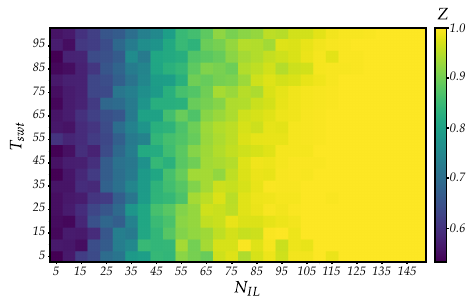}}
\caption{Time-averaged interlayer order parameter $Z$ (colour bar) as a function
of the switching period $T_{swt}$ and the number of interlayer links $N_{IL}$,
averaged over 30 network realisations ($T_{swt}$ in steps of $5$ from $5$ to
$100$; $N_{IL}$ in steps of $5$ from $5$ to $150$). Bright (yellow) marks
near-complete interlayer synchronisation, $Z\to 1$. The crossover to $Z\simeq 1$
shifts to larger $N_{IL}$ as $T_{swt}$ increases, tracking---and lying just above,
for the reason discussed in Section~\ref{sec:TLE}---the $\lambda_\perp=0$
stability boundary of Fig.~\ref{fig:TLE}.}
\label{fig:Z_Tswt_NIL}
\end{figure}

The complete picture is shown in Fig.~\ref{fig:Z_Tswt_NIL}, which plots $Z$ as a
function of both $T_{swt}$ and $N_{IL}$, sweeping $T_{swt}$ in steps of~5 from~5
to~100 and $N_{IL}$ in steps of~5 from~5 to~150, averaged over 30 network
realisations.

The order-parameter analysis establishes that periodic redistribution of the
interlayer links strongly promotes interlayer synchronisation: near-complete
synchronisation (which we define throughout as $Z\gtrsim 0.95$, i.e.
$1-Z\lesssim 0.05$) is achieved with $N_{IL} \approx N/3\approx 100$ active links
under switching, compared to the requirement $N_{IL} = N$ in the static case. As
quantified in Section~\ref{sec:TLE}, the order parameter becomes indistinguishable
from unity ($1-Z\to 0$) at a slightly larger link number, $N_{IL}\approx 120$--$130$.

\section{Master Stability Analysis}
\label{sec:MSF}

To support the order-parameter analysis, we carry out a master stability function
(MSF) analysis. We first give a brief derivation of the MSF for the present
system, following the interlayer master-stability approach for identical multiplex
layers~\cite{sevilla2016,tang2019}, specialised to the nonlocally coupled
Kuramoto--Sakaguchi ring; an analogous transverse-stability analysis for a finite
Kuramoto--Sakaguchi array is given in Ref.~\cite{lee2021chimera}.

From the definition of the interlayer order parameter \eqref{eq:R12}, the
synchronisation manifold is
\[
\theta_{i}^{(1)}(t)=\theta_{i}^{(2)}(t)= s_{i}(t) \qquad \forall\, i \in \{1,2,\dots,N\}.
\]
On this manifold, the dynamics of each node are governed by
\begin{equation}
\dot{s}_{i}(t) = \omega_{i} - \frac{\sigma}{2R}\sum_{j=i-R}^{i+R} \sin\!\big(s_{i}-s_{j}+\alpha\big).
\label{eq:syncstate}
\end{equation}
On the interlayer synchronisation manifold, the two interlayer coupling terms
reduce to the common value $\sin(\alpha_{12})$, which is identical for both
replicas; the manifold $\theta_i^{(1)}=\theta_i^{(2)}$ is therefore invariant for
any $\alpha_{12}$. For $\alpha_{12}=0$ the interlayer terms vanish and the
synchronised motion is governed by Eq.~\eqref{eq:syncstate}; for
$\alpha_{12}\neq 0$ each active replica pair additionally acquires the common drive
$\chi_i(t)\,\sigma_{12}\sin\alpha_{12}$, which shifts the base trajectory but,
being identical in both layers, cancels in the interlayer difference and hence does
not affect the transverse stability analysis below. To analyse the stability of the
synchronised state, we introduce small perturbations
\begin{eqnarray}
\theta_{i}^{(1)}(t) &=& s_{i}(t)+ \delta \xi_{i}^{(1)}(t), \label{eq:theta1perturb}\\
\theta_{i}^{(2)}(t) &=& s_{i}(t)+ \delta \xi_{i}^{(2)}(t), \label{eq:theta2perturb}
\end{eqnarray}
where $\delta\ll 1$ and $\xi_{i}^{(\ell)}(t)$ are the perturbations of layer
$\ell \in \{1,2\}$. Substituting \eqref{eq:theta1perturb}--\eqref{eq:theta2perturb}
into \eqref{eq:model1}--\eqref{eq:model2} and expanding to first order in $\delta$,
with $C_{ij}(t)=\cos(s_i-s_j+\alpha)$ and the instantaneous interlayer coupling of
replica pair $(i,i)$ written as $\tilde{\sigma}_i(t)=\chi_i(t)\,\sigma_{12}$, gives
\begin{eqnarray}
\dot{\xi}_i^{(1)} &=& -\frac{\sigma}{2R}\sum_{j=i-R}^{i+R} C_{ij}(t)\left(\xi_i^{(1)} - \xi_j^{(1)}\right) \nonumber \\
& & + \tilde{\sigma}_i(t)\cos\alpha_{12} \left(\xi_i^{(2)} - \xi_i^{(1)}\right), \label{eq:xi1dot}\\
\dot{\xi}_i^{(2)} &= &-\frac{\sigma}{2R}\sum_{j=i-R}^{i+R} C_{ij}(t)\left(\xi_i^{(2)} - \xi_j^{(2)}\right) \nonumber\\
&& + \tilde{\sigma}_i(t)\cos\alpha_{12} \left(\xi_i^{(1)} - \xi_i^{(2)}\right). \label{eq:xi2dot}
\end{eqnarray}

Introducing the interlayer difference perturbation
\[
\eta_{i}(t) = \xi_{i}^{(1)}(t)-\xi_{i}^{(2)}(t),
\]
and subtracting \eqref{eq:xi2dot} from \eqref{eq:xi1dot} gives
\begin{multline}
\dot{\eta}_{i}(t) = -\frac{\sigma}{2R} \sum_{j=i-R}^{i+R} C_{ij}(t) (\eta_{i}-\eta_{j}) - 2 \tilde{\sigma}_i(t)\cos\alpha_{12}\, \eta_{i} \\= -\sum_{j=1}^{N} L_{ij}(t) \eta_{j}-2\tilde{\sigma}_i(t)\cos\alpha_{12}\, \eta_{i},
\end{multline}
with the time-dependent weighted Laplacian
\begin{equation}
L_{ij}(t) = \begin{cases}
-\dfrac{\sigma}{2R} C_{ij}(t), & \text{if } |i-j| \leq R \text{ and } i \neq j, \\[4pt]
\dfrac{\sigma}{2R}\displaystyle\sum_{k=i-R}^{i+R} C_{ik}(t), & \text{if } i = j, \\[4pt]
0, & \text{otherwise},
\end{cases}
\end{equation}
and the diagonal interlayer-coupling matrix
\begin{equation}
\boldsymbol{\Sigma}_{12}(t) = \operatorname{diag}\!\big(\tilde{\sigma}_1(t),\tilde{\sigma}_2(t),\dots, \tilde{\sigma}_N(t)\big).
\label{eq:Sigma12}
\end{equation}
Collecting the perturbations into the vector
$\boldsymbol{\eta}(t)=(\eta_1,\dots,\eta_N)^\top$, the transverse dynamics take
the compact form
\begin{equation}
\dot{\boldsymbol{\eta}}(t) = \Bigl[-\mathbf{L}(\mathbf{s}(t)) - 2\cos\alpha_{12}\,\boldsymbol{\Sigma}_{12}(t)\Bigr]\boldsymbol{\eta}(t),
\label{eq:eta_compact_MSF}
\end{equation}
which is the equation whose largest Lyapunov exponent is evaluated in
Section~\ref{sec:TLE}. The intralayer Laplacian term $-\mathbf{L}$ is diffusive,
and the interlayer term $-2\cos\alpha_{12}\,\boldsymbol{\Sigma}_{12}$ is
dissipative whenever $\sigma_{12}\cos\alpha_{12}>0$, so that active interlayer
links contract transverse perturbations. For $\alpha_{12}=0$, which is used
throughout, the interlayer term reduces to $-2\boldsymbol{\Sigma}_{12}(t)$.

We define the \emph{master stability function} $\Psi(\sigma_{12})$ as the largest
Lyapunov exponent of Eq.~\eqref{eq:eta_compact_MSF} evaluated numerically as a
function of the interlayer coupling strength $\sigma_{12}$, with all other
parameters fixed. In the switching case, $\Psi$ also depends on $T_{swt}$, and we
write $\Psi(\sigma_{12}, T_{swt})$. Interlayer synchronisation is linearly stable
when $\Psi < 0$.

It should be stressed that $\Psi$ and the finite-time transverse Lyapunov
exponent $\lambda_\perp$ studied in Section~\ref{sec:TLE} are \emph{the same
object}: both are the largest Lyapunov exponent of the transverse variational
system~\eqref{eq:eta_compact_MSF}, computed by the same numerical scheme. The two
notations distinguish the two complementary parameter cuts we use to map the
stability boundary. In the present section we sweep the coupling strength and
display $\Psi(\sigma_{12},T_{swt})$ at fixed link number, which is the natural MSF
viewpoint and connects directly to the classical coupling-strength stability
curves of Pecora and Carroll~\cite{pecora1998}. In Section~\ref{sec:TLE} we
instead fix the physical coupling strength at its working value
$\sigma_{12}=0.01$ and map the same exponent, written $\lambda_\perp$, over the
$(N_{IL},T_{swt})$ plane, which is the natural viewpoint for the design question of
how many links and how fast a switching are required. Reporting both cuts of a
single exponent makes the internal consistency of the analysis explicit and
guards against the impression that two independent stability criteria are being
invoked.

\begin{figure}
\centerline{\includegraphics[width=\columnwidth]{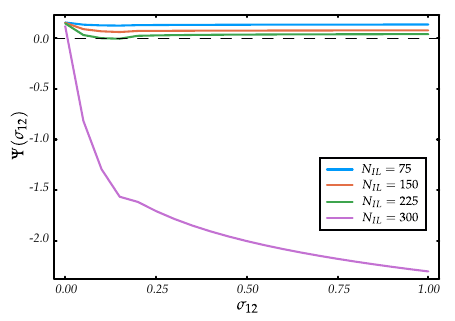}}
\caption{Master stability function $\Psi(\sigma_{12})$ for $N=300$ in the static
case, for $N_{IL}=75$ (blue), $N_{IL}=150$ (orange), $N_{IL}=225$ (green),
and $N_{IL}=300$ (purple). The dashed line marks $\Psi=0$. Only $N_{IL}=300$
yields $\Psi<0$.}
\label{fig:MSFStatic}
\end{figure}

We first study the static case. Figure~\ref{fig:MSFStatic} shows
$\Psi(\sigma_{12})$ for $N_{IL} = 75$, $150$, $225$, and $300$. Only for
$N_{IL}=300$ does the MSF become negative, independent of $\sigma_{12}$; for
all partial sets of static links, $\Psi > 0$ throughout the range $\sigma_{12}\in[0,1]$.
This is fully consistent with the order-parameter result that full synchronisation
requires $N_{IL} = N$ in the static case.

\begin{figure}
\centerline{\includegraphics[width=\columnwidth]{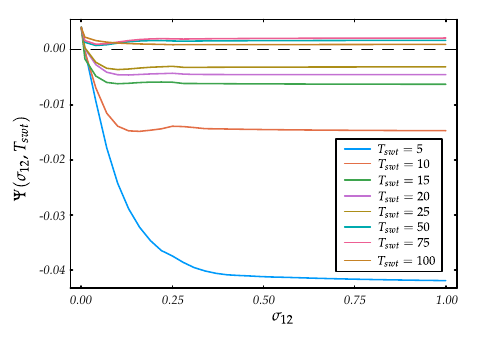}}
\caption{Master stability function $\Psi(\sigma_{12},T_{swt})$ for $N=300$,
$N_{IL}=75$, and switching times $T_{swt}\in\{5,10,15,20,25,50,75,100\}$
(see legend). The dashed line marks $\Psi=0$. Short switching periods
($T_{swt}\leq 25$) yield $\Psi<0$; longer periods remain unstable.}
\label{fig:MSFswitching}
\end{figure}

We next compute the MSF for the switching case, fixing $N_{IL}=75$ and varying
$T_{swt}\in\{5,10,15,20,25,50,75,100\}$. Figure~\ref{fig:MSFswitching} shows
that $\Psi(\sigma_{12},T_{swt}) < 0$ for $T_{swt}=5,10,15,20,25$, while
$\Psi > 0$ for $T_{swt}=50,75,100$. Short switching periods therefore stabilise
the transverse direction at a link number ($N_{IL}=75$) where static coupling
fails entirely; longer periods are insufficient.

\begin{figure}
\centerline{\includegraphics[width=\columnwidth]{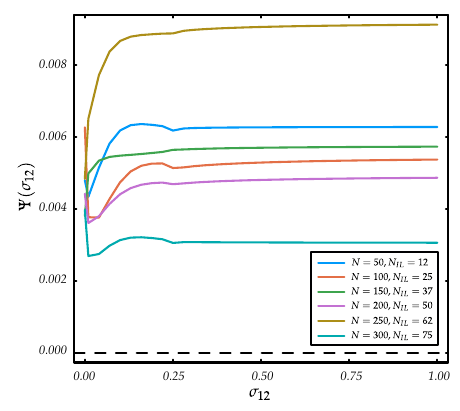}}
\caption{Master stability function $\Psi(\sigma_{12})$ for fixed interlayer link
density $N_{IL}/N=0.25$ in the static case, for network sizes
$N=50$ ($N_{IL}=12$), $100$ ($N_{IL}=25$), $150$ ($N_{IL}=37$),
$200$ ($N_{IL}=50$), $250$ ($N_{IL}=62$), and $300$ ($N_{IL}=75$).
$\Psi>0$ in all cases; static 25\% link density cannot synchronise any
of the network sizes tested.}
\label{fig:MSFstatiPopulation}
\end{figure}

We also examine how network size affects synchronisation. Fixing the interlayer
link density $N_{IL}/N = 0.25$ and varying $N\in\{50,100,150,200,250,300\}$
(with $N_{IL} = \lfloor N/4 \rfloor$), Fig.~\ref{fig:MSFstatiPopulation} shows
that in the static case $\Psi > 0$ for every network size: a 25\% link density is
insufficient for static synchronisation regardless of $N$.

\begin{figure}
\centerline{\includegraphics[width=\columnwidth]{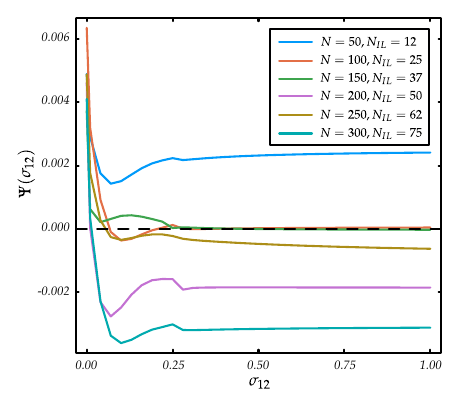}}
\caption{Master stability function $\Psi(\sigma_{12})$ for fixed interlayer link
density $N_{IL}/N=0.25$ and $T_{swt}=25$, for the same network sizes as
Fig.~\ref{fig:MSFstatiPopulation}. Under switching, $\Psi<0$ for
$N \geq 200$; $N=150$ is marginally stable; $N \leq 100$ remain unstable.}
\label{fig:MSFswtPopulation}
\end{figure}

Under switching with $T_{swt}=25$ and the same link density,
Fig.~\ref{fig:MSFswtPopulation} shows a clear size dependence: for
$N \geq 200$, $\Psi$ is clearly negative, indicating stable interlayer
synchronisation. For $N=150$, $\Psi$ is marginally negative above a
threshold in $\sigma_{12}$, while for $N \leq 100$ the MSF remains positive.
Larger networks thus benefit more from switching at fixed relative link density.

\section{Transverse Lyapunov Exponent}
\label{sec:TLE}

\subsection{Definition and stability criterion}

The master stability analysis of Section~\ref{sec:MSF} yields the linearised
equation governing the growth or decay of interlayer perturbations
$\boldsymbol{\eta}(t)=(\eta_1,\dots,\eta_N)^\top$. The long-time averaged
exponential rate of change of $\|\boldsymbol{\eta}\|$ is quantified by the
\emph{transverse Lyapunov exponent} (TLE)~\cite{fujisaka1983,pecora1998}

\begin{equation}
\lambda_{\perp} \;=\; \lim_{T\to\infty} \frac{1}{T}\ln\frac{\|\boldsymbol{\eta}(T)\|}{\|\boldsymbol{\eta}(0)\|}.
\label{eq:TLE_def}
\end{equation}

$\lambda_{\perp}$ measures the largest Lyapunov exponent restricted to the subspace
\emph{transverse} to the interlayer synchronisation manifold
$\theta_i^{(1)}=\theta_i^{(2)}$. The stability criterion is:
$\lambda_{\perp} < 0$ if and only if the interlayer synchronisation manifold is
exponentially stable. When $\lambda_\perp > 0$, infinitesimal perturbations grow on
average and the synchronised state is unstable; the boundary $\lambda_\perp = 0$
marks the critical number of interlayer links $N_{IL}^{*}$ at which the basin of
attraction of the desynchronised state collapses.

\subsection{Linearised transverse dynamics}

The closed evolution equation for $\boldsymbol{\eta}$, derived in
Section~\ref{sec:MSF} [Eq.~\eqref{eq:eta_compact_MSF}], reads in component form

\begin{equation}
\dot{\eta}_{i}(t)
  = -\frac{\sigma}{2R}\!\sum_{\substack{j=i-R \\ j\neq i}}^{i+R}\!
        C_{ij}(t)\!\left(\eta_{i}-\eta_{j}\right)
    - 2\,\tilde{\sigma}_i(t)\cos\alpha_{12}\,\eta_{i},
\label{eq:eta_evol}
\end{equation}

where $C_{ij}(t)=\cos\!\bigl(s_i(t)-s_j(t)+\alpha\bigr)$ is evaluated on the
chimera base state $\mathbf{s}(t)$, periodic-boundary sums are understood ($j$
taken modulo $N$), and $\tilde{\sigma}_i(t)=\chi_i(t)\sigma_{12}\in\{0,\sigma_{12}\}$
is the time-varying interlayer coupling of node $i$. In compact matrix form,

\begin{equation}
\dot{\boldsymbol{\eta}}
  = \Bigl[-\mathbf{L}\!\bigl(\mathbf{s}(t)\bigr)
          - 2\cos\alpha_{12}\,\boldsymbol{\Sigma}_{12}(t)\Bigr]\boldsymbol{\eta},
\label{eq:eta_matrix}
\end{equation}

where $\mathbf{L}(\mathbf{s}(t))$ is the weighted graph Laplacian of the
intralayer nonlocal ring (with time-dependent weights $C_{ij}$) and
$\boldsymbol{\Sigma}_{12}(t)=\mathrm{diag}(\tilde\sigma_1(t),\dots,\tilde\sigma_N(t))$.
The interlayer phase lag enters only through the scalar factor $\cos\alpha_{12}$;
with $\sigma_{12}\cos\alpha_{12}>0$, the term $-2\tilde\sigma_i(t)\cos\alpha_{12}\,\eta_i$
is stabilising: it drives perturbations toward zero whenever node $i$ participates
in an interlayer link. For $\alpha_{12}=0$, $\cos\alpha_{12}=1$. The Laplacian
term $-\mathbf{L}\boldsymbol{\eta}$ is generally not sign-definite because
$C_{ij}$ changes sign across the coherent and incoherent domains of the chimera.

\subsection{Numerical computation: Benettin--Galgani algorithm}

Because $\mathbf{s}(t)$ is a chimera state (neither a fixed point nor a periodic
orbit in the large-$N$ limit), the coefficient matrix in \eqref{eq:eta_matrix}
cannot be diagonalised analytically and $\lambda_\perp$ must be estimated
numerically. We employ the single-vector Benettin--Galgani renormalisation
algorithm~\cite{benettin1980}:

\begin{enumerate}
\item \textbf{Transient.} Starting from random initial phases, integrate the base-state
equation~\eqref{eq:syncstate} for $T_{\mathrm{tr}}=1000$ time units using a
fourth-order Runge--Kutta scheme with $\mathrm{d}t=0.01$ to allow the chimera to
settle onto its attracting state.

\item \textbf{Initialise perturbation.} Draw a random unit vector
$\boldsymbol{\eta}(0)\in\mathbb{R}^N$  from the standard Gaussian
distribution and normalise it.

\item \textbf{Simultaneous integration.} Advance both $\mathbf{s}(t)$ and
$\boldsymbol{\eta}(t)$ by one RK4 step using the coupled system
\eqref{eq:syncstate}--\eqref{eq:eta_evol}.

\item \textbf{Renormalise and accumulate.} After each step $k$, record the
stretch factor $\varrho_k = \|\boldsymbol{\eta}(t_k)\|$, renormalise
$\boldsymbol{\eta}\leftarrow\boldsymbol{\eta}/\varrho_k$, and accumulate
$\ln\varrho_k$.

\item \textbf{Average.} After a measurement window of length $T_{\mathrm{meas}}$,
\begin{equation}
\lambda_{\perp}
  \;\approx\;
  \frac{1}{T_{\mathrm{meas}}}
  \sum_{k=1}^{T_{\mathrm{meas}}/\mathrm{d}t} \ln\varrho_k.
\label{eq:TLE_numerical}
\end{equation}
\end{enumerate}

The length of the measurement window is the single most important convergence
parameter in the switching case, because $\lambda_\perp$ must average over the
nonautonomous coupling and not merely over a few switching events. We therefore
adapt the window to the switching period, using
$T_{\mathrm{meas}} = \max(2000,\,200\,T_{swt})$, which guarantees at least
$200$ complete switching cycles per measurement at every $T_{swt}$ sampled (up to
$T_{\mathrm{meas}}=2\times 10^4$ for $T_{swt}=100$). This adaptive rule is what
makes the finite-time exponent a faithful estimate of the asymptotic transverse
growth rate of the \emph{switched} system rather than an artefact of a particular
short sequence of link configurations. For the time-switching topology, the
active-node set $\mathcal{A}(t)\subset\{1,\dots,N\}$ with $|\mathcal{A}|=N_{IL}$
is redrawn uniformly at random at the start of each interval of length $T_{swt}$, 
and the identical switching
schedule is applied to both the base-state and the variational integration, so
that the variational dynamics see exactly the coupling that the trajectory
experiences. The static case is recovered by drawing $\mathcal{A}$ once and
holding it fixed for the entire window. Fixing all three random seeds (initial
condition, perturbation, and switching sequence) makes every reported value
exactly reproducible.

\subsection{Results}

\begin{figure}
\centerline{\includegraphics[width=\columnwidth]{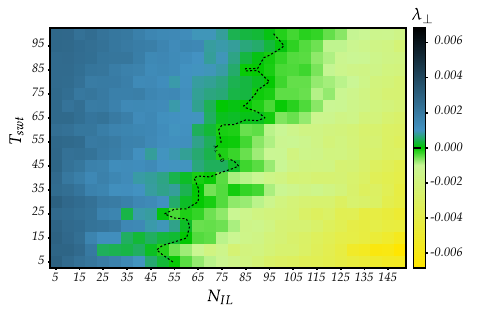}}
\caption{Transverse Lyapunov exponent $\lambda_\perp$ as a function of $N_{IL}$
and $T_{swt}$ for $N=300$. The dashed contour marks $\lambda_\perp=0$; the
yellow region ($\lambda_\perp<0$) indicates stable interlayer synchronisation
and the blue region ($\lambda_\perp>0$) indicates transverse instability.}
\label{fig:TLE}
\end{figure}

Figure~\ref{fig:TLE} shows $\lambda_\perp$ as a joint function of $N_{IL}$ and
$T_{swt}$ for $N=300$. The dashed contour marks $\lambda_\perp=0$, separating
the stable regime (yellow, $\lambda_\perp<0$, to the right) from the unstable
regime (blue, $\lambda_\perp>0$, to the left). For all switching times, there
exists a critical number $N_{IL}^*$ beyond which the interlayer synchronous
manifold is attracting. This critical number decreases as $T_{swt}$ decreases:
faster switching requires fewer interlayer links to achieve transverse stability.
For short switching periods ($T_{swt}\lesssim 25$), $N_{IL}^* \approx 50$--$65$,
whereas for longer periods ($T_{swt}\gtrsim 75$), $N_{IL}^*$ rises to
approximately~100. These results are fully consistent with the order-parameter
analysis of Section~\ref{sec:order} and the MSF analysis of
Section~\ref{sec:MSF}, and confirm that temporal redistribution of a fixed, sparse
set of interlayer links can stabilise replica-node coherence in the chimera regime.

The stability map in Fig.~\ref{fig:TLE} also closes the loop with the other two
diagnostics quantitatively. The $\lambda_\perp=0$ contour is monotonic and
single-valued in both arguments: for every switching period there is a unique
critical link number $N_{IL}^*(T_{swt})$, and $N_{IL}^*$ increases smoothly with
$T_{swt}$ with no reentrant islands of stability or instability. This monotonicity
is the signature of a single mechanism---temporal spreading of the coupling
resource---rather than of resonances between the switching period and any internal
time scale of the chimera. The boundary runs parallel to, and slightly below (in
$N_{IL}$), the $Z\to 1$ crossover of Fig.~\ref{fig:Z_Tswt_NIL} (the offset is
quantified below), and its fast-switching corner ($T_{swt}\lesssim 25$,
$N_{IL}^*\approx 50$--$65$) reproduces the sign change of the MSF curves in
Fig.~\ref{fig:MSFswitching}. The three diagnostics thus agree along the whole
transition curve---the strongest consistency check available short of the
analytical reduction that the heterogeneous chimeric base state precludes.

\begin{figure}
\centerline{\includegraphics[width=\columnwidth]{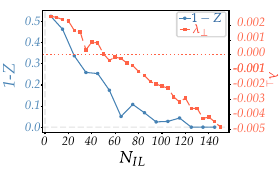}}
\caption{Residual interlayer incoherence $1-Z$ (blue, left axis) and transverse
Lyapunov exponent $\lambda_\perp$ (red, right axis) as functions of the number of
interlayer links $N_{IL}$, at fixed switching period $T_{swt}=25$ ($N=300$,
$\sigma_{12}=0.01$). Both indicators decrease monotonically with $N_{IL}$. The
horizontal dotted line marks $\lambda_\perp=0$: the manifold becomes linearly
attracting ($\lambda_\perp<0$) at $N_{IL}^*\approx 55$--$65$, where the residual
incoherence $1-Z\approx 0.3$ is still appreciable. Near-complete synchronisation
($Z\gtrsim 0.95$) is reached at $N_{IL}\approx N/3\approx 100$, and the order
parameter saturates ($1-Z\to 0$) only near $N_{IL}\approx 120$--$130$. The offset
between the stability onset and the coherence thresholds reflects the small
contraction rate just above threshold, where switching-re-excited relaxation keeps
the time-averaged mismatch finite even though $\lambda_\perp<0$ (see text).}
\label{fig:gap}
\end{figure}

It is worth making explicit how the linear-stability threshold relates to the
saturation of the order parameter, because the two are not numerically identical
and the distinction could otherwise be mistaken for an inconsistency.
Figure~\ref{fig:gap} resolves the point directly by overlaying the two
diagnostics, $1-Z$ and $\lambda_\perp$, as functions of $N_{IL}$ at fixed
$T_{swt}=25$. Both decrease monotonically with $N_{IL}$, confirming that they
diagnose the same underlying transition, but they cross their respective
reference levels at different link numbers. The two diagnostics thus define a consistent
ladder of three link numbers at $T_{swt}=25$. The transverse Lyapunov exponent
changes sign at the linear-stability onset $N_{IL}^*\approx 55$--$65$, where $1-Z$
is still appreciable ($\approx 0.3$); the near-complete-synchronisation threshold
$Z\gtrsim 0.95$ used as the headline criterion in Section~\ref{sec:order} is
reached at $N_{IL}\approx N/3\approx 100$; and the residual interlayer incoherence
$1-Z$ does not fall to near zero until $N_{IL}\approx 120$--$130$. The condition
$\lambda_\perp<0$ thus marks the onset of \emph{linear} attractivity of the
interlayer manifold, whereas $Z\gtrsim 0.95$ and, ultimately, $1-Z\to 0$ require
the progressively stronger, quantitative condition that the observed replica-phase
mismatch be small.

There is no contradiction between the two, and the reason is precisely why $1-Z$
remains nonzero while $\lambda_\perp<0$. A negative $\lambda_\perp$ guarantees
that an infinitesimal transverse perturbation decays \emph{asymptotically}, at the
mean rate $|\lambda_\perp|$, on the invariant interlayer manifold. It does not
guarantee that the \emph{observed}, finite-amplitude mismatch vanishes, for three
reasons that all operate just above threshold. First, immediately above
$N_{IL}^*$ the contraction rate $|\lambda_\perp|$ is small (Fig.~\ref{fig:gap},
right axis: $|\lambda_\perp|\sim 10^{-3}$ near the crossing), so the transverse
relaxation time $1/|\lambda_\perp|$ is of the order of $10^3$ time units, often
longer than several switching intervals. Second, the coupling is genuinely
time-varying: every switching event at period $T_{swt}$ abruptly changes which
nodes are contracted, re-injecting energy into the transverse subspace and
preventing the mismatch from ever reaching the asymptotic zero that the sign of
$\lambda_\perp$ alone would suggest. The system therefore settles into a
statistically stationary state with a small but nonzero residual mismatch, whose
size is set by the balance between the switching ``kicks'' and the contraction
rate---hence $1-Z\propto 1/|\lambda_\perp|$ to leading order, large near threshold
and decaying as $|\lambda_\perp|$ grows. Third, $Z$ is an average over a finite
window and finite $N$, so even a perfectly attracting manifold retains
$O(N^{-1/2})$ fluctuations. As $N_{IL}$ increases past $N_{IL}^*$,
$|\lambda_\perp|$ grows, the transverse relaxation becomes fast compared with
$T_{swt}$, the switching kicks are damped between events, and $1-Z\to 0$. The
linear-stability boundary is therefore a lower bound on the link number required
for visibly tight locking, and the gap between $N_{IL}^*$ and the $Z\simeq1$
threshold visible in Fig.~\ref{fig:gap} is itself a measure of how strongly
attracting the manifold is: a wide gap signals weak attraction near onset, a
narrow one strong attraction.

A remark on convergence is in order. Because nonlocally coupled chimeras at finite
$N$ are long chaotic transients~\cite{wolfrum2011}, the finite-time exponent is
the physically meaningful quantity, and its value depends on the measurement
window. We verified that the sign of $\lambda_\perp$---the only feature on which
the stability classification depends---is stable under doubling of
$T_{\mathrm{meas}}$ for grid points away from the contour, and that the residual
scatter near $\lambda_\perp=0$ is confined to a band one grid cell wide, which we
take as the practical uncertainty on the location of $N_{IL}^*$. The adaptive
window $T_{\mathrm{meas}}=\max(2000,200\,T_{swt})$ introduced in
Section~\ref{sec:TLE} is essential here: a fixed short window would
systematically misclassify the large-$T_{swt}$ rows, where only a handful of
switching events would otherwise be sampled.

\section{Conclusion}

We have studied interlayer synchronisation in a duplex of two identical,
nonlocally coupled Kuramoto--Sakaguchi rings, each in the chimera regime and joined
by a sparse, weak, time-varying set of symmetric interlayer links. Three consistent
diagnostics---the time-averaged order parameter $Z$, the master stability function
$\Psi(\sigma_{12},T_{swt})$, and the transverse Lyapunov exponent
$\lambda_\perp$---characterise the transition as a function of the link number
$N_{IL}$ and the switching period $T_{swt}$. The central finding is that temporal
switching renders the interlayer-synchronous manifold attracting with far fewer
links than static coupling requires: the static case needs all-to-all coupling
($N_{IL}=N=300$, with $\Psi>0$ for every $N_{IL}<N$), whereas switching with
$T_{swt}\le 25$ reaches near-complete synchronisation ($Z\gtrsim 0.95$) at
$N_{IL}\approx N/3\approx 100$ and has $\Psi<0$ already at $N_{IL}=75$. The
transverse-Lyapunov map (Fig.~\ref{fig:TLE}) gives the full $(N_{IL},T_{swt})$
phase diagram, with the critical number $N_{IL}^*$ falling as switching speeds up.

The mechanism is the temporal redistribution of the coupling resource: at any
instant only $N_{IL}$ pairs are linked, but switching samples all replica pairs
over time, spreading the stabilising effect without raising the instantaneous link
count---consistent, in the fast-switching limit, with the effective-coupling
picture of blinking-network theory~\cite{belykh2004,stilwell2006}. Establishing
this in the chimera regime required a transverse-stability analysis valid when the
synchronous trajectory is spatially heterogeneous: because the chimeric base state
makes the variational operator~\eqref{eq:eta_compact_MSF} non-co-diagonalisable
with a single Laplacian, the eigenmode reduction of the multiplex MSF~\cite{tang2019}
does not apply, and we integrate the full transverse system instead. The same
exponent, shown either as $\Psi(\sigma_{12},T_{swt})$ or as
$\lambda_\perp(N_{IL},T_{swt})$, is one internally consistent diagnostic, and its
agreement with $Z$ along the entire transition confirms that the synchronisation is
genuinely attracting rather than a coincidental resemblance of two patterns.

A clear finite-size effect appears: at fixed link density $N_{IL}/N=0.25$ and
$T_{swt}=25$, switching stabilises the manifold for $N\ge 200$ but not for smaller
networks, because the temporal spreading of coupling is more effective when more
nodes are available over which to redistribute the links.

Throughout, the intralayer chimera is preserved---coherent and incoherent domains
survive and the layers never collapse into global coherence---so replica-node
coherence and partial intralayer coherence are compatible. This is what makes the
model a faithful caricature of the two cerebral hemispheres joined by costly, gated
callosal projections~\cite{anesiadis2022,ramlow2019}: the layers lock as replicas
\emph{while keeping} their internal chimera, as unihemispheric sleep requires, and
the temporal link redistribution mirrors dynamic functional connectivity. The
opposite event---collapse of the chimera into global coherence, which our analysis
identifies as loss of the partially coherent manifold rather than a change in
synchronisability---echoes the proposed analogy between chimera collapse and
epileptic-seizure onset~\cite{andrzejak2016}. These correspondences are structural
rather than literal, but they indicate where the mechanism may be relevant.

Several limitations bound these conclusions. The layers are identical and the
oscillators share $\omega_i=0$, which isolates the roles of switching and phase-lag
frustration but excludes frequency disorder and structural layer mismatch.
Finite-$N$ chimeras are long chaotic transients~\cite{wolfrum2011}, so we report a
finite-time exponent over a window matched to the switching period; the stability
verdict is that of the chimeric state actually observed. Finally, the
effective-coupling picture is rigorous only in the fast-switching
limit~\cite{belykh2004,stilwell2006}, so the genuinely nonautonomous behaviour at
larger $T_{swt}$---the failure of stabilisation for $T_{swt}\ge 50$ at
$N_{IL}=75$---lies outside its validity.

These limitations set a natural agenda: to locate, in the $(N_{IL},T_{swt})$ plane,
where the fast-switching approximation ceases to predict the measured
$\lambda_\perp$; to map $N_{IL}^*$ against the chimera-controlling parameters
$\alpha$ and $r=R/N$; to test layers supporting mismatched chimera patterns,
coupling ranges, or heterogeneous frequencies; to replace uniform random
reselection with structured or adaptive switching rules; and to examine finite-size
scaling of the boundary toward the thermodynamic limit. Each direction extends
interlayer-synchronisation theory further into the spatially heterogeneous,
partially coherent regime that the chimera state exemplifies.

\section*{Author Declarations}

\subsection*{Conflict of Interest}
The authors have no conflicts to disclose.

\subsection*{Author Contributions}
Both authors conceived the study and developed the model. M.~C.~Eser implemented
the numerical simulations and the transverse-stability computations. M.~Riza
supervised the work. Both authors analysed the results and wrote the manuscript.

\section*{Data Availability}

The data and the simulation and analysis code that support the findings of this
study are available from the corresponding author upon reasonable request.

\bibliographystyle{aipnum4-2}
\bibliography{references}

\end{document}